\documentclass{aa}

\usepackage{natbib}

\usepackage[inline]{enumitem}
\usepackage{amsmath}
\usepackage{xcolor}
\usepackage{graphicx}
\graphicspath{{images/}}
\usepackage{txfonts}

\usepackage[normalem]{ulem}

\begin{document}

   \title{Chirp mass--distance distributions \\ of the sources  of gravitational waves}
   \author{M. Ossowski
          }

   \institute{Institute of Theoretical Physics, Faculty of Physics,
University of Warsaw,
              Pasteura 5, 02-093 Warsaw, Poland\\
              \email{maciej.ossowski@fuw.edu.pl}
             }

   \date{Received 25 September 2019 ; accepted 26 January 2021}
  \abstract
   {The detection of gravitational waves emitted by binary black holes raises questions regarding the  origin of binaries. There are several models in the literature involving binary evolution in both the field and clusters.}
   {We compare the predictions of these models with observations and establish the reliability of this comparison.}
   {We use the likelihood calculation to compare the models in the space spanned by the observed chirp mass and the luminosity distance of the source.}
   {We rank the models by their ability to explain all current gravitational wave detections. It is shown that the most probable models correspond to binary evolution with low metallicity. Several variants of such evolution have similar likelihoods. The globular cluster model, considered here, is disfavoured.}
   {We present the usefulness of the method in distinguishing between models when new observations become available. We calculate the number of observations required to distinguish between each pair of models.  We find that the number varies from ten to several thousand for some pairs of models, yet almost two-thirds of pairs are distinguishable with at most 100 observations.}


   \maketitle
%________________________________________________________________

\section{Introduction}
The discovery of gravitational waves (GWs;\citealp{PhysRevX.6.041014}) marked the beginning of gravitational wave astronomy. 
The recent catalogue of gravitational wave sources \citep{2018GWCatalogarXiv181112907T,LIGO3} contains 50 candidates, 47 of which are binary black hole (BBH) mergers.
The properties of the population were studied by \cite{2017MNRAS.471.4702B}.
In the mean time, several models of formation of BBHs have been proposed.
The leading model is based on the assumption that all these sources form as a result of binary evolution.
This idea has been explored by several authors in recent decades \citep{1997astro.ph..1134L,2002ApJ...572..407B,2008ApJS..174..223B,2012ApJ...759...52D}.
Recently a version of this scenario in which the stars undergo strong mixing and have uniform chemical composition was studied as an alternative option.
Additionally, several authors discuss the possibility of enhanced formation of BBHs in globular clusters \citep{2016PhRvD..93h4029R,2017MNRAS.464L..36A}.

The availability of data for 47 BBH mergers allows the model predictions to be compared with observations.
We present these comparisons using Bayesian statistical inference.
The paper is organised as follows: In Sect. 2 we present the principles of detection of BBHs, and Sect. 3 contains a description of the statistical method used.
In Sect. 4 we present the models of BBH populations used for comparison and in Sect. 5 we discuss the properties of the populations that are important for this analysis.
In Sect. 6 we briefly explain why signal-to-noise ratio distributions carry little information about the population that is useful for the Advanced Laser Interferometer Gravitational-Wave Observatory (LIGO), and finally in Sect. 7 we present results of the analysis of the chirp mass--luminosity distance distributions in the context of current detections and the possibility of distinguishing between models when more data become available.
\section{Detection in binary systems}
Consider a BBH with masses $m_1$ and $m_2$, which is moving in a circular orbit.
\cite{1963PhRv..131..435P} showed that those systems will emit energy via GWs.
The waveform of these GWs  depends on the redshifted {chirp mass} $\mathcal{M}_z=\mathcal{M}(1+z)$, where chirp mass is defined as
\begin{displaymath}
\mathcal{M}=\frac{(m_1m_2)^{3/5}}{(m_1+m_2)^{1/5}}. 
\end{displaymath} 
The position of the BBH on the sky with respect to the detector is described in spherical coordinates by angles $\theta$ and $\phi$.
The orientation of the BBH is described by the angles $\iota$ and $\psi$ in spherical coordinates, where the z-axis is parallel to the total angular momentum of the system \citep{Finn1996PhRvD..53.2878F}.
Measuring distance to the system must take the geometry of spacetime into account.
These cosmological effects are encompassed by the luminosity distance
\begin{displaymath}
D_L(z)=(1+z)D_H \int_{0}^{z}\dfrac{dz^ \prime}{E(z^\prime)},
\end{displaymath}
\noindent
where $E(z)=\sqrt{\Omega_m(1+z)^3+\Omega_\Lambda}$ and $\ D_H=c / H_0$.
We use the following parameters:
$\Omega_m = 0.27$, $\Omega_\Lambda = 1-\Omega_m =0.73$, and $H_0=70.4 \, \mathrm{ km }\, s^{-1}\,\mathrm{ Mpc}^{-1}$.
We assume that the BBH will be detected if its signal-to-noise ratio (S/N) $\rho$ is greater than 8.
The S/N here is defined as
\begin{equation}
\label{eq:SNR}
\rho=\frac{A(\mathcal{M}_z)\Theta(\theta, \phi, \iota, \psi) }{D_L},
\end{equation}  
where $A(\mathcal{M}_z)$ 
is the luminosity distance at which an optimally oriented  binary with redshifted chirp mass $\mathcal{M}_z$ is detected with an S/N of four
and $\Theta(\theta, \phi, \iota, \psi)$ is the  sensitivity of the detector to the orientation of the  BBH. 
The orientation sensitivity $\Theta \in[0,4]$ is 
\begin{displaymath}
\Theta(\theta, \phi, \iota, \psi)= 2[F^2_+(1+\cos^2\iota)^2+4F^2_\times \cos^2 \iota]^{1/2},
\end{displaymath}
where $F_+$ and $F_\times$ are the sensitivity of the  detector to polarisations $+$ and $\times$ of the GW \citep{Finn1996PhRvD..53.2878F} defined as
   \begin{eqnarray}
      F_+=\frac{1}{2}(1+\cos^2\theta)\cos2\phi \cos2\psi-\cos\theta \sin2\phi \sin2\psi, \\
      F_\times=\frac{1}{2}(1+\cos^2\theta)\cos2\phi \sin2\psi+\cos\theta \sin2\phi \cos2\psi .
   \end{eqnarray}

\begin{figure}[ht]
   \centering
   \includegraphics[width=\hsize]{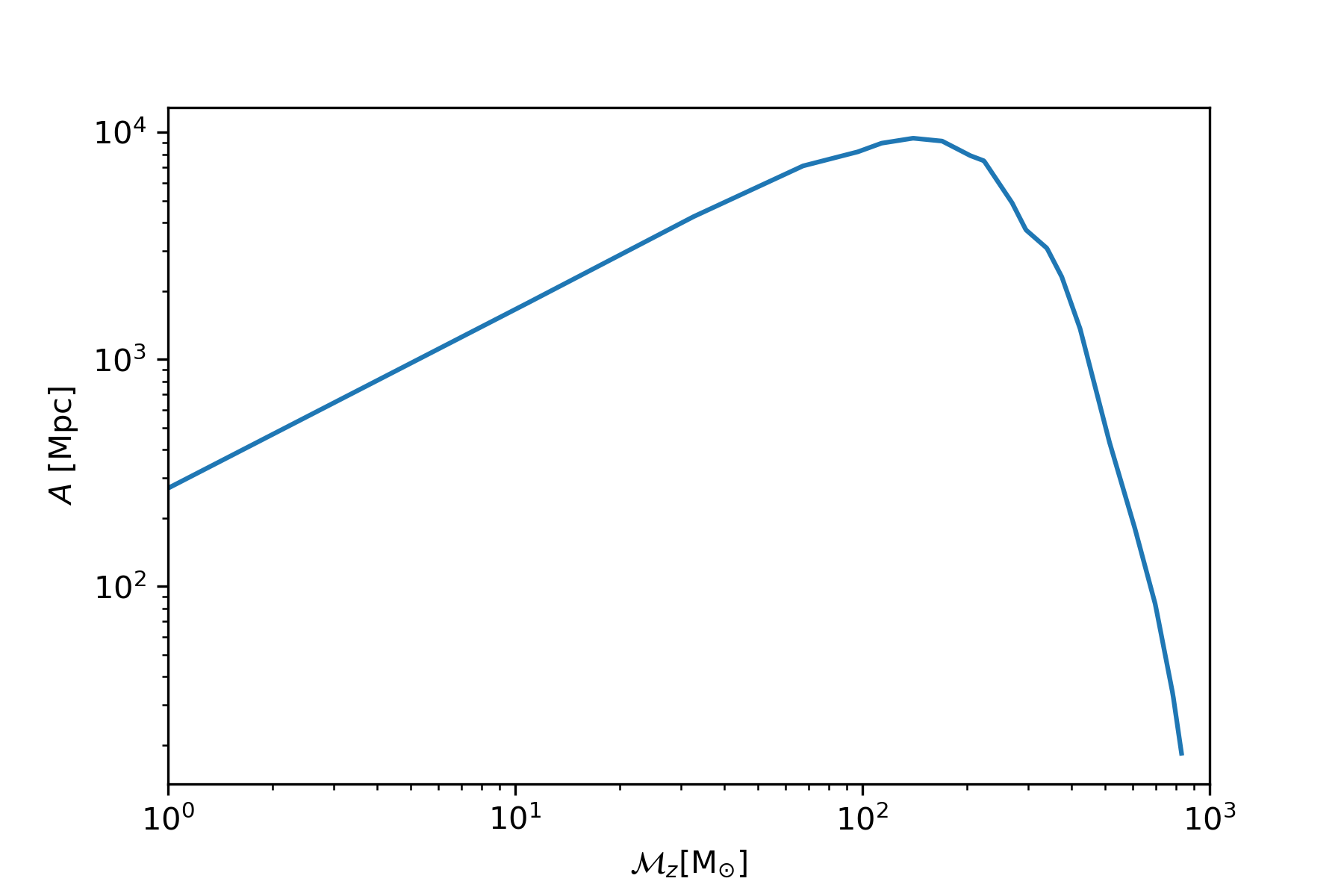}
      \caption{Response of Advanced LIGO  to the observed chirp mass.}
         \label{fig:detH1}
   \end{figure}

We consider the Advanced LIGO in configuration H1 \citep{2016PhRvD..93k2004M}. The purpose of this choice is twofold: \begin{enumerate*}[label=(\roman*)]
    \item to determine which model is preferred by up-to-date observations and
    \item to determine if this detector configuration is capable of differentiating between models.
\end{enumerate*}
For $\mathcal{M}_z \leq 100 M_{\odot}$, the dependence of the  sensitivity of Advanced LIGO  on chirp mass can be approximated by ${A(\mathcal{M}_z) \propto  \mathcal{M}_z^{5/6}}$, which leads to the commonly used approximation of S/N ${\rho \propto  \mathcal{M}_z^{5/6} \Theta/D_L}$.
The full range of $A(\mathcal{M}_z)$, which is twice the horizon luminosity distance (i.e. distance for which $\rho=8$ for a given detector), is presented in Fig.~\ref{fig:detH1}. 

The dependence of sensitivity on chirp mass essentially encapsulates the ability of the interferometer  to detect different frequencies which is a function of detector construction and corresponding noises.
The main sources of the noise are seismic noises directly moving interferometer mirrors, thermal noise interacting with the beam, and quantum noise producing fluctuations in the number of photons in the beam.
A full discussion on the noise sources can be found in \cite{2016PhRvD..93k2004M}.

\section{Method of distinguishing models}
To distinguish between the models we employ the method of statistical Bayesian inference.
A broader description of this approach can be found in \cite{bayesian}.
For the purpose of this analysis let us define:

\begin{tabular}{lp{0.7\linewidth}}
     $P(M_i \mid O)$ & probability that model $M_i$ is correct, given observation $O$ \\
     $P(O \mid M_i)$ & probability of observing $O$, given that model $M_i$ is correct\\
     $P(M_i)$ & prior probability that model $M_i$ is correct.
\end{tabular}

We note that
\begin{equation}
P (O \mid M_i)=\mathcal{L}(O \mid M_i):=\prod_{k\in O }\frac{dP_i}{d\mathcal{B}}(\mathcal{B}_k),
\label{eq:likelihood}
\end{equation}
where $\mathcal{L}(O \mid M_i)$ is the likelihood function for the model $M_i$, $\mathcal{B}$ is the set of investigated observables, and $k$ indexes the observations in $\mathcal{O}$.
Furthermore, we denote $O$ as $O_i$ to emphasise that is was generated using model $M_i$.
The probability distribution of $\mathcal{B}=(D_L,\mathcal{M})$ for a given model is defined  as follows:
\begin{equation}
  \frac{dP_i}{  d\mathcal{B}} (\mathcal{B})=\sum_l \delta(D_L-D_{Ll})\delta(\mathcal{M}-\mathcal{M}_l),
\end{equation}
where for a given model $M_i$ we estimate the above probability by calculating a set of $10^5$ observations labelled by $l$ and characterised by $(D_{Ll},\mathcal{M}_l)$.
The delta function is approximated using the value of the nearest bin on a 200 by 200 grid.
Subsequently, for a model $M_i$ we can calculate the likelihood of this model given up-to-date, real GW observations by
\begin{equation}
    \mathcal{L}_{i\text{,Real}}=\prod_n \frac{dP_i}{d\mathcal{B}} (D_L = D_{Ln},\mathcal{M}=\mathcal{M}_n),
    \label{eq:likelihoodReadObs}
\end{equation}
where $n$ covers the set of the real GW observations.
Assuming that we are in the position of maximal ignorance with regard to the validity of the model, we let all $P(M_i)$ be equal. Then, following \cite{bayesian}, let us define the Bayes factor as
\begin{equation}
    F_{ij}:=\frac{P(M_i \mid O_i)}{P(M_j \mid O_i)}=\frac{P(O_i \mid M_i)}{P(O_i \mid M_j)},
\end{equation}
where the second equality is a result of the Bayes theorem,
    \begin{displaymath}
    P(M_i \mid O_i)=\frac{P(O_i \mid M_i)P(M_i)}{P(O_i)},
    \end{displaymath}
where $P(O_i)$ is the normalisation.

We would like to point out that $F_{ij} \gg 1$ means that observation $O_i$ is in favour of model $M_i$ over $M_j$ and thus suggests that the former better represents reality than the latter.
We analyse distributions of Bayes factors for different models to establish the threshold of distinguishability with p-value $\alpha$. 
The {threshold of distinguishability} (TOD) with p-value $\alpha$ is defined as the minimal number of observations needed for $(1 -\alpha) \times 100\% $ of the Bayes factors $F_{ij}$ to be greater than 1.
To avoid numerical noise we also require that the same inequality holds for every number of the observations greater than the threshold.
The process of calculating the TOD is as follows:
\begin{enumerate}
\item Generate an observation $O_i$ using model $M_i$.
\item For each pair of models, calculate $F_{ij}$ using the number of observations ranging logarithmically from 3 to 3\,000.
\item Repeat the previous point $10^4$ times if the number of observations in $O_i$ is smaller than 10 or $10^3$ times otherwise, then analyse the distributions of the Bayes Factor based on this sample.
\end{enumerate}

We note the following characteristics of this approach:
\begin{itemize}
\item There is no symmetry in reversing the models $F_{ij}\neq F_{ji} ^{-1}$. This is because those Bayes factors use observations based on different models.
\item The mean value of $F_{ij}$ is expected to be greater than 1. Similarly to the  previous point, this is is because on average obtaining $O_i$ is more probable with $M_i$ than with $M_j$. 
This stresses the importance of using a high enough p-value.
\item The TOD is dependent on the choice of the detector and the set of observables $\mathcal{B}$.

\end{itemize}

\section{Investigated models}
We surveyed a total of 34 models for compatibility with up-to-date observations. Additionally, 16 out of the above 34 models were surveyed for distinguishability based on their ability to explain observations.
Of those 34 models, 32 are \texttt{StarTrack} models available at \texttt{syntheticuniverse.org}. To minimise numerical error it was additionally required that a model have at least 100 systems evolved to BBH.
We now present an abridged description of the models; for the full description of the \texttt{StarTrack} models we refer to \cite{2012ApJ...759...52D}. The two non-\texttt{StarTrack} models are chemically homogeneous evolution and evolution in globular clusters.

\subsection{\texttt{StarTrack} models}
\texttt{StarTrack} is a population synthesis code focused on evolution of field stars and binary systems \citep{2002ApJ...572..407B,2008ApJS..174..223B}. It incorporates a twofold treatment of the Hertzsprung Gap donors in a common envelope (CE) phase. In the family of models B, the CE phase initiated by a Hertzsprung Gap donor always leads to
a merger and formation of a Throne-Żytkow object. These models usually do not form a sufficient number of BBHs to explain observations.

In the family of models A, the energy balance is calculated following the \cite{1984ApJ...277..355WWebbink} formalism as
\begin{equation}
\bigg( \frac{GM_{\mathrm{don,f}M_{\mathrm{acc}}}}{2A_{\mathrm{f}}}-\frac{GM_{\mathrm{don,i}M_{\mathrm{acc}}}}{2A_{\mathrm{i}}}\bigg)=\frac{GM_{\mathrm{don,i}}M_\mathrm{{don,env}} }{\lambda R_{\mathrm{don,lob}}},
\end{equation}
where $i$ and $f$ stand for the initial and final, and $M_{\mathrm{don}}$, $M_{\mathrm{acc}}$, and $M_\mathrm{{don,env}}$ are the masses of the donor, the accretor, and the ejected envelope. $R_{\mathrm{don,lob}}$ is the Roche lobe radius of the donor at the 
beginning of the Roche Lobe Overflow (ROFL) and $A$ is the binary separation. The energy balance determines whether the binary will shed its envelope, merge during CE, or coalesce. The $\lambda$ parameter is used in energy balance of CE and is defined as
\begin{equation}
E_{\mathrm{bind}}=-\frac{GM_{\mathrm{don}}M_\mathrm{{don,env}} }{\lambda R},
\end{equation}
where $E_{\mathrm{bind}}$ is the binding energy of the envelope and $R$ is the radius of the donor.

The compact object  merger can  only occur if the binary survives the CE phase or phases. Therefore, we consider only the family of models A. \texttt{StarTrack} models can be categorised according to differences in the input physics: CE energy parameter $\lambda$, maximum neutron star (NS) mass, natal kick velocity, supernova engine, wind loss, and mass transfer treatment; for a more detailed discussion see \citep{2012ApJ...759...52D}.We used the model files published at the website \texttt{syntheticuniverse.org} and the description below refers to the models provided on that site.

\subsubsection{Standard model}
Standard model uses the $\lambda$ parameter varying during evolution according to \cite{0004-637X-716-1-114XuLi}, which is dubbed as \textit{Nanjing} $\lambda$. Maximal NS mass is taken to be $2.5 \mathrm{M_{\odot}}$. The mass transfer is half-conservative (half of the mass is ejected to infinity). This means that $f_a=0.5$, defined as $\dot{M}_{\mathrm{acc}}=f_a\dot{M}_{\mathrm{don}}$, where $\dot{M}_{\mathrm{acc}}$ is ROFL mass transfer rate and $\dot{M}_{\mathrm{don}}$ is the ROFL mass transfer rate of the  donor. Natal kick distribution is Maxwellian with $\sigma=265 \mathrm{km\ s^{-1}}$ and the fall-back factor $f_{fb}$ is according to \cite{2012ApJ...749...91FFryer}. The fall-back factor $f_{fb}$ is defined as $v_k=v_{\mathrm{max}}(1-f_{fb})$, where $v_{\mathrm{max}}$ is velocity drawn from the Maxwellian distribution and $v_k$ is the final velocity. The Standard model uses a Rapid supernova engine, as described in \cite{2012ApJ...749...91FFryer}.
\subsubsection{Variants}
In Table \ref{tab:variations} we present variants of the \texttt{StarTrack} model. Parameters other than those mentioned under `Changed parameter' remain the same as in the Standard model.

\begin{table}[ht]
           % title of Table
     % is used to refer this table in the text
\centering   
\caption{Variants of \texttt{StarTrack} model}% used for centering table
\begin{tabular}{c c c c}        % centered columns (4 columns)
\hline\hline                 % inserts double horizontal lines
Model & Changed parameter  \\    % table heading 
\hline                        % inserts single horizontal line
   V1 & fixed $\lambda=0.01$  \\      % inserting body of the table
   V2 & fixed $\lambda=0.1$  \\
   V3 & fixed $\lambda=1$  \\
   V4 & fixed $\lambda=10$  \\
   V5 & $M_{\mathrm{NS,max}}=3.0 \mathrm{M_{\odot}}$  \\ 
   V6 & $M_{\mathrm{NS,max}}=2.0 \mathrm{M_{\odot}}$  \\ 
   V7 & natal kick distribution's $\sigma = 132.5 \mathrm{km / s}$  \\ 
   V8 & full BH natal kicks $f_{fb}=1$\\ 
   V9 & no BH natal kicks $f_{fb}=0$\\
   V10 &  delayed SN engine \citep{2012ApJ...749...91FFryer}\\ 
   V11 & wind mass loss reduced to $50\%$  \\ 
   V12 & fully conservative mass transfer $f_a=1$\\ 
   V13 & fully non-conservative mass transfer $f_a=0$\\ 
   V14 & \textit{Nanjing} $\lambda$ increased by factor 5  \\ 
   V15 & \textit{Nanjing} $\lambda$ decreased by factor 5  \\ 
   
\hline                                   %inserts single line
\end{tabular}
\label{tab:variations} 
\tablefoot{ Except for the stated change, parameters are the same as in the Standard model}
% \tablebib{(1) \cite{2012ApJ...759...52D}}.
\end{table}

\subsection{Other models}
\begin{enumerate}
\item Chemically homogeneous evolution (CHE): One of the alternative channels of BBH evolution is CHE, described in \cite{2016MNRAS.460.3545D}. Here we study the model denoted `Default' in \cite{2016MNRAS.458.2634M}. Owing to rapid rotation causing transport of matter between the core and the surface, CHE systems do not have a chemical gradient. Such stars gradually shrink, staying inside the Roche Lobe, and omit a transfer phase. Because of their non-violent evolution, the resulting BBHs may have a higher mass than is typical for these objects.
\item Globular Clusters (GCs): \cite{2013ApJ...763L..15M} suggest that the merger rate in GCs may be comparable to the rate in the field evolution. Because of this we include a GC model simulated with MOCCA code \citep{2013MNRAS.431.2184G} and broadly described in \cite{2017MNRAS.464L..36A}. This model is a sum of approximately 2000 submodels with varying initial conditions: $Z \in [0.002 - 0.02]$ and distance from the galaxy centre between $1$ and $50$ kpc. All submodels share the Maxwellian natal velocity kick distribution with $\sigma = 265 \mathrm{km \ s^{-1}}$, the rotational velocity of a cluster of $220\mathrm{km\ s^{-1}}$, and the {IMF} according to \cite{2001MNRAS.322..231K}. The simulated GCs agree with observations of GCs in the Galaxy \citep{2017MNRAS.464L..36A}. 
Coalescing in a GC leads to faster BBH evolution, while those expelled and coalescing outside are subject to slower evolution; both scenarios are considered.
\end{enumerate}

\section{Characteristics of BBH population}
We consider a space of three (not entirely independent) observables: \{$\rho$, $\mathcal{M}$, $D_L$\}. We also show in Sect. \ref{sec:luminosity} that the luminosity distributions contain little information, and therefore we limit our analysis to two-dimensional chirp mass--luminosity distance distributions.
For the purpose of this analysis, the BBH models have only two characteristic attributes: intrinsic chirp mass distribution and spatial distribution. 

We have \citep{2012PhRvD..85b3535T}
\begin{equation}
\label{eq:dNdzdThdM}
\frac{d^4N}{dt d\Theta dz d \mathcal{M}}=\frac{dV_c}{dz}\frac{\dot{n}(z)}{1+z} \mathcal{P}(\mathcal{M})\mathcal{P}(\Theta) ,
\end{equation}
where $\dot{n}(z)$ is the merger rate density containing information about the rate of merging of the BBH at a given redshift and $\mathcal{P}(\mathcal{M})$ is the observed chirp mass distribution (in contrast to the intrinsic chirp mass distribution used by \citeauthor{2012PhRvD..85b3535T}). If $D_c(z)=D_H \int_{0}^{z}\dfrac{dz^ \prime}{E(z^\prime)}$ is comoving distance and $V_c$ is comoving volume, then
\begin{displaymath}
\frac{dV_c}{dz}= \frac{4\pi D_c(z)^2 D_H}{E(z)}.
\end{displaymath}
Integrating Eq. (\ref{eq:dNdzdThdM}) over $\Theta$, neglecting the time dependence, and substituting the variable $z$ for $D_L$ we get
\begin{equation}
\frac{d^2N}{dD_L d \mathcal{M}}=\frac{4\pi D_c^2(z)D_H}{E(z)D_c(z)+D_H(1+z)}\frac{\dot{n}(z)}{1+z} \mathcal{P}(\mathcal{M}),
\end{equation}
which we use as $P(O\mid M_i)$, setting $\mathcal{B}$ to be $\{D_L, \mathcal{M}\}$.

\begin{figure}[ht]
   \centering
   \includegraphics[width=\hsize]{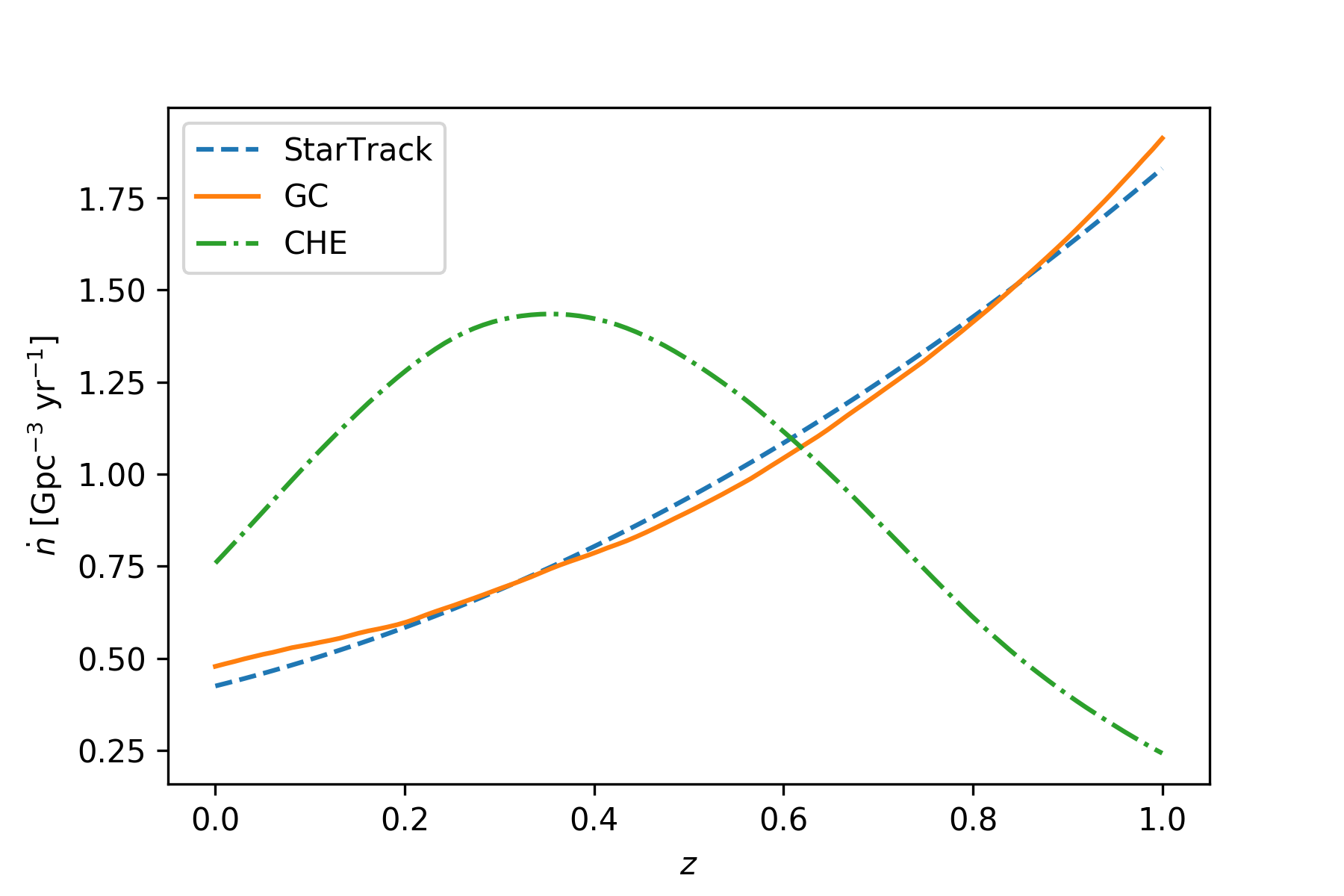}
      \caption{Normalised merger rate distributions of the used models. The dashed, solid, and dot-dashed lines correspond to \texttt{StarTrack}, GC, and CHE models, respectively. \texttt{StarTrack} merger rate density is taken from model M10 from \citet{2017MNRAS.471.4702B}}
         \label{fig:MRD}
   \end{figure}

Alternatively, integrating over $\mathcal{M}$ and $\Theta$ we get
\begin{equation}
\frac{dN}{dz}=\frac{4\pi D_c(z)^2 D_H}{E(z)}\frac{\dot{n}(z)}{1+z},
\label{eq:dNdz}
\end{equation}
which is, when normalised, the probability of observing an event originating in the coalescence at the redshift $z$ (regardless of its S/N).
To choose the normalisation we limit redshift to $z_{max}$, which is the solution of equation $\frac{A (\mathcal{M}_z)\Theta_{max}}{D_L(z)}=\rho_{min}$ for the greatest mass in the model.  Figure \ref{fig:MRD} presents the distributions of merger rate density. Because $\dot{n}$ enters Eq. (\ref{eq:dNdz}) linearly, and this equation itself is normalised, the absolute magnitude of merger rate density is unimportant. 

\begin{figure}[ht]
   \centering
   \includegraphics[width=\hsize]{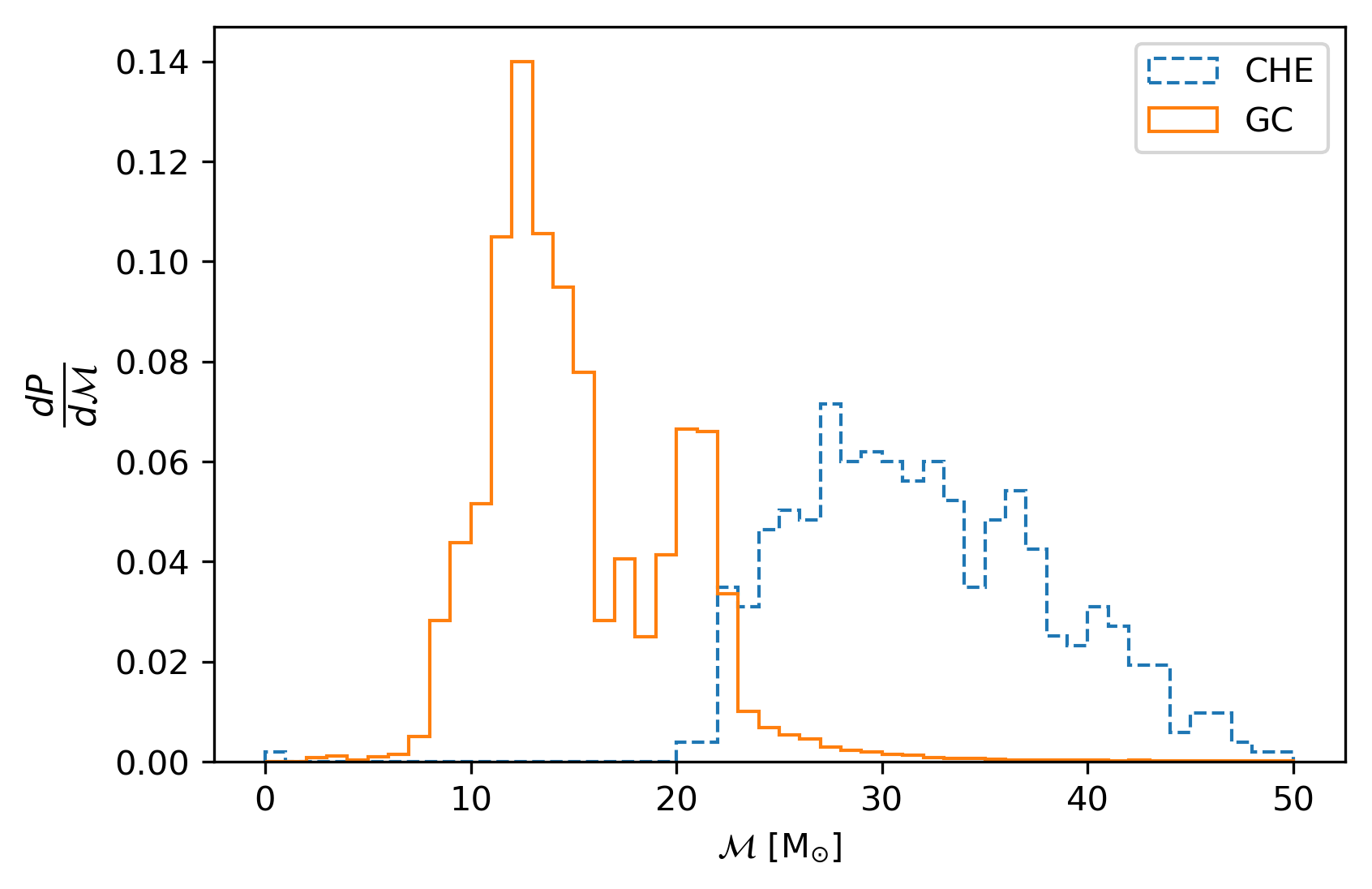}
      \caption{Normalised intrinsic chirp mass distributions. The solid line corresponds to the GC model and the dashed line corresponds to the CHE model.}
         \label{fig:mass}
   \end{figure}

In Fig. \ref{fig:mass} we present the chirp mass distributions of CHE and GC models. The distributions for the \texttt{StarTrack} models can be found in \cite{2012ApJ...759...52D}; they typically average around $7 M_{\odot}$ for $Z=1 Z_{\odot}$ and under $14 M_{\odot}$ for $Z=0.1 Z_{\odot}$. 
   
\section{Luminosity distribution}
\label{sec:luminosity}
Before investigating the mass--distance distribution, let us first consider a luminosity distribution. Because of the  short range of Advanced LIGO (compared to other third-generation detectors, e.g. the Einstein Telescope) we can approximate the luminosity distance $D_L$ with a physical distance $R$. In Eq. (\ref{eq:SNR}), we have $R\propto \rho^{-1}$. Then, assuming a uniform distribution of sources and a single value of mass, we have $N\propto V \propto R^3 \propto \rho^{-3}$ and $\frac{dN}{d\rho}\propto \rho^{-4}$. This raises a concern that, unless the BBH spatial distributions are sufficiently different, all S/N distributions may show little deviation from $\rho^{-4}$. 

\begin{figure}[ht]
  \centering
  \includegraphics[width=\hsize]{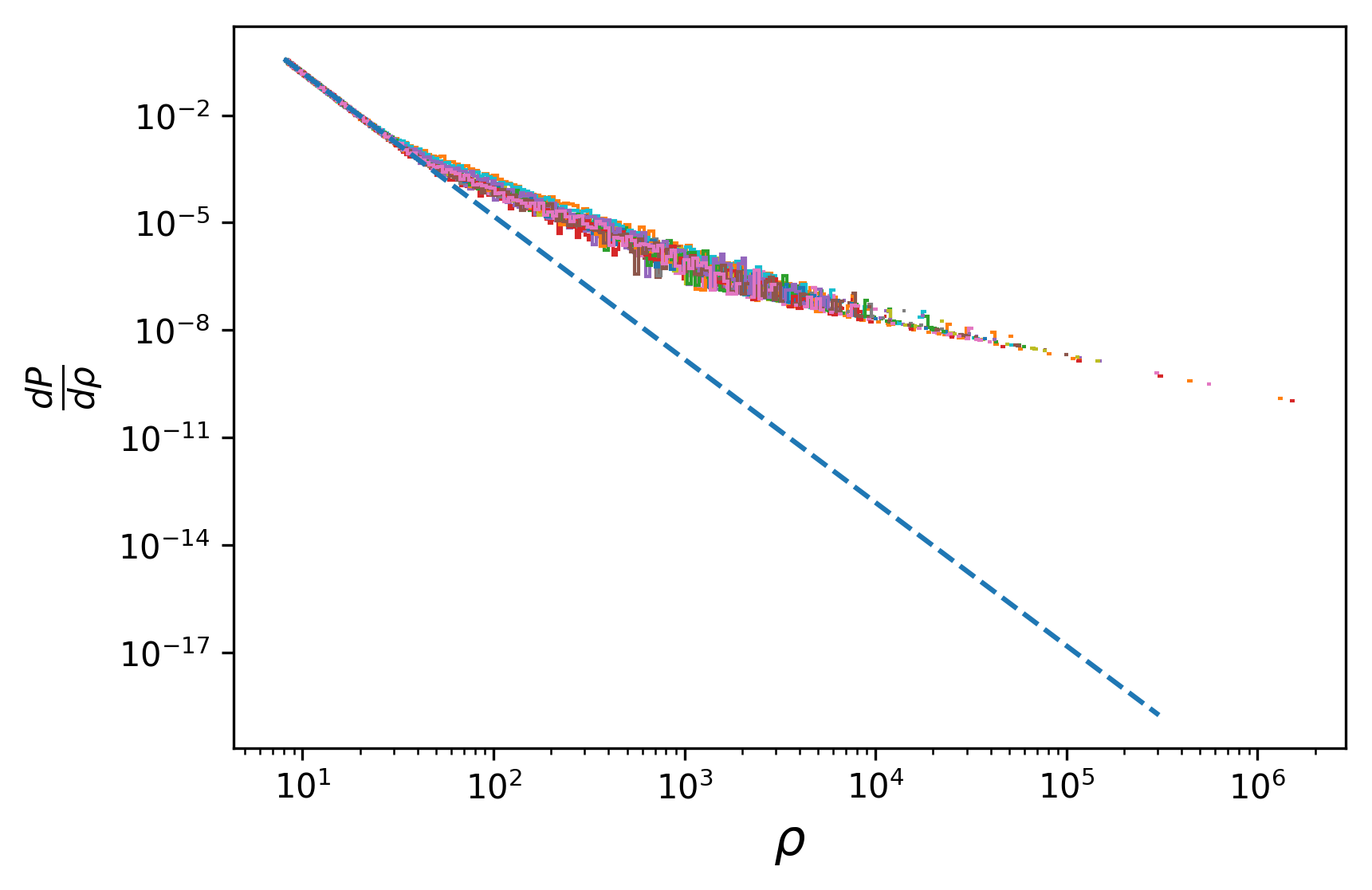}
      \caption{Normalised S/N distribution. The dashed line represents the theoretical curve, with $\rho_{min}=8$. Other curves are the distributions of 16 simulated models (listed in Sect. \ref{ssec:likelihood}) based on $10^5$ binaries.}
         \label{fig:SNR}
  \end{figure}

We demonstrate in Fig. \ref{fig:SNR} that for the majority of the events the distributions follow the theoretical curve, with only the most luminous events showing a difference. For this reason, we investigated the distributions of chirp mass--luminosity distance instead of S/N. The difference between the theoretical curve and calculations in the case of the 
most luminous events with $\rho> 10^2$ is mostly due to the mass distribution of merging BBHs.
The heavy BBHs contribute to the tail with high $\rho$ if the events are nearby, while for smaller values of $\rho$ the S/N distribution is averaged over the large volume where events are detectable.

\section{Results}
\subsection{Likelihoods}
 The updated list of GW observations can be found in the catalogues by \cite{2018GWCatalogarXiv181112907T} and \cite{LIGO3} These observations are used in the calculation of the likelihood of explaining the observations by any single model (\ref{eq:likelihoodReadObs}). The likelihoods are calculated according to Eq. \ref{eq:likelihood} with the chirp mass--luminosity distance taken as $P(D \mid M_i)$. In the calculation of probabilities, the errors are accounted for by weighting the probabilities with Gaussian distributions corresponding to the measurement uncertainty of the given observations.

\begin{figure*}
   \centering
   \includegraphics[width=\textwidth]{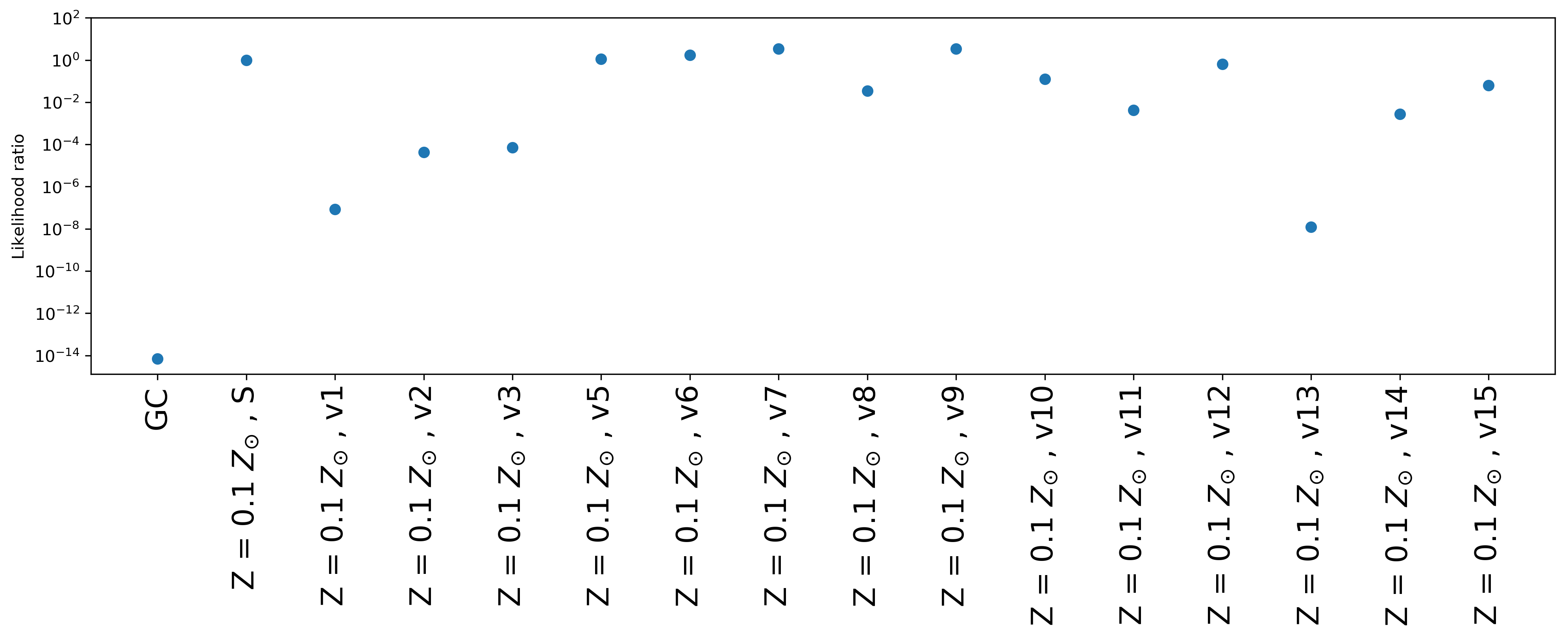}

   \caption{Likelihoods for the 16 selected models that could explain all of the selected GW observations. The likelihoods are normalised by the likelihood of the Standard StarTrack model.}
              \label{fig:like}%
    \end{figure*}

Selected models with the highest likelihood are subsequently examined for distinguishability using Bayesian inference in Sect. \ref{ssec:thesholds}. We present those likelihoods in  Fig. \ref{fig:like}.
The demarcation point is the natural one; the likelihood of the next most likely model when compared to the likelihood of the Standard model is
of the order of the numerical error.

All of the solar metallicity \texttt{StarTrack} models could explain only the closest observations with the smallest mass, with chirp mass of several solar masses and distance of less then $800 Mpc$. On the contrary, the chirp mass distribution of the CHE model  is too high to explain the mergers with chirp mass lower than $20 M_\odot$. The \texttt{StarTrack} models with $Z=0.1Z_{\odot}$ (with exception of the Variant 4 whose intrinsic mass distribution is too small) and GC are selected for calculations of thresholds. It is worth emphasising that none of the models are capable of explaining observations further than $3 \mathrm{Gpc}$.

Variant 9, characterised by no BH natal kick, is the most probable model with likelihood $\approx3.5$ times higher than the Standard model.
Several models are slightly preferred to the Standard model: V5, V6, V7, and V9.
Those models differ from the  Standard model in that they have either a higher or lower NS mass and lower or no natal kick, respectively.
Still none of these likelihoods are sufficiently high for the model to be a meaningful alternative to the Standard model.
Any change to $\lambda$ decreases the likelihood of the model by a factor of at least $100$.
An increase in wind mass loss (V11) reduces the likelihood by 8 orders of magnitude.
Finally, the observations strongly disfavour the GC model, as it is less likely by 15 orders of magnitude.

\label{ssec:likelihood}
\subsection{Thresholds of distinguishability}
For the calculation of the TOD, the data $O_i$ are taken to have Gaussian errors with 
\begin{displaymath}
\frac{\sigma_{D_L}}{D_L}=\genfrac{}{}{0pt}{0}{+0.384}{-0.318}
\end{displaymath}
\begin{displaymath}
\frac{\sigma_{\mathcal{M}}}{\mathcal{M}}=\genfrac{}{}{0pt}{0}{+0.112}{-0.077},
\end{displaymath}
which are average relative errors for all real observations. It should be noted that because the catalogues give a $90\%$ symmetric interval error, the above errors were scaled to correspond to a standard deviation. Although the errors cited by the catalogues are not Gaussian, they differ from Gaussian mainly at the tails of the distribution and thus approximation is appropriate.
The purpose of calculating the thresholds is to check whether likelihoods in the Fig. \ref{fig:like} are reliable with probability $\alpha$, given the available number of observations.
Table \ref{tab:perc10} suggests that two groups of parameters can be distinguished (i.e. thresholds for all pairs are under 47) with the current number of observations: wind mass reduction (Standard, V11) and $\lambda=\text{const}$ (V1, V2, V3, also for the purpose of the comparison Standard). Comparing this result with Fig. \ref{fig:like} we can conclude that changes to the Standard model parameters are unpreferable. Some groups of parameters are distinguishable, except for distinguishing with the Standard model. In this group there are: BH natal kick (V8, V9), mass transfer (V12, V13),  and Nanjing $\lambda$ (V14, V15). If for other reasons the Standard model would not be preferable, the results for those groups point to the likely correction of the parameters. The remaining groups: maximal NS mass (V5, V6), natal kick distribution (V7), and SN engine (V10) are not distinguishable. Finally, the GC, V11, and V13 models are (separately) distinguishable from the rest of the models. The overall situation may change if research based on other evidence limits the available model space.

\begin{figure}[ht]
   \centering
   \includegraphics[width=\hsize]{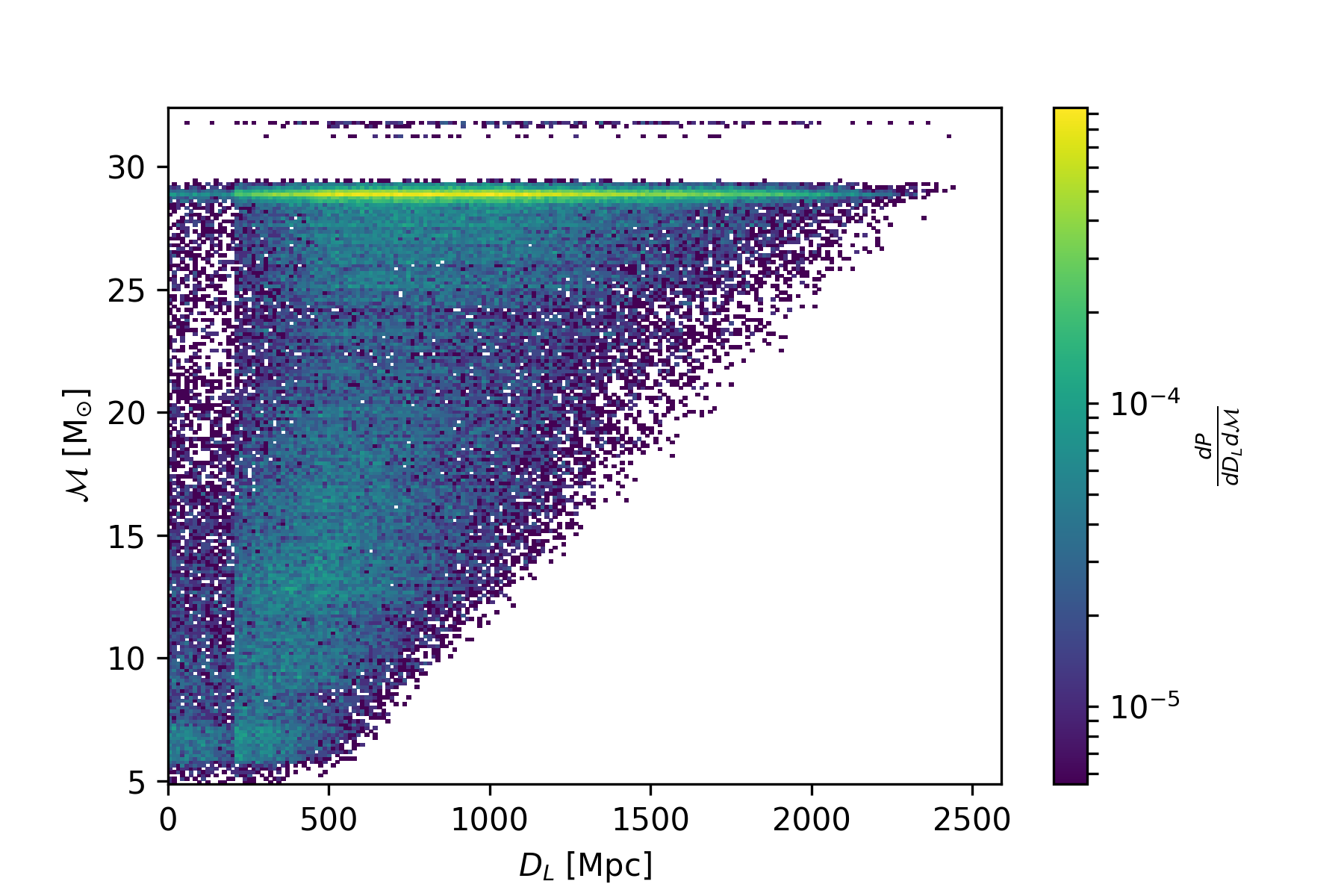}
      \caption{Chirp mass--luminosity distance distribution for the Standard \texttt{StarTrack} model with $Z=0.1Z_{\odot}$.}
         \label{fig:ABHBH.002S}
   \end{figure}

\begin{figure}[ht]
   \centering
   \includegraphics[width=\hsize]{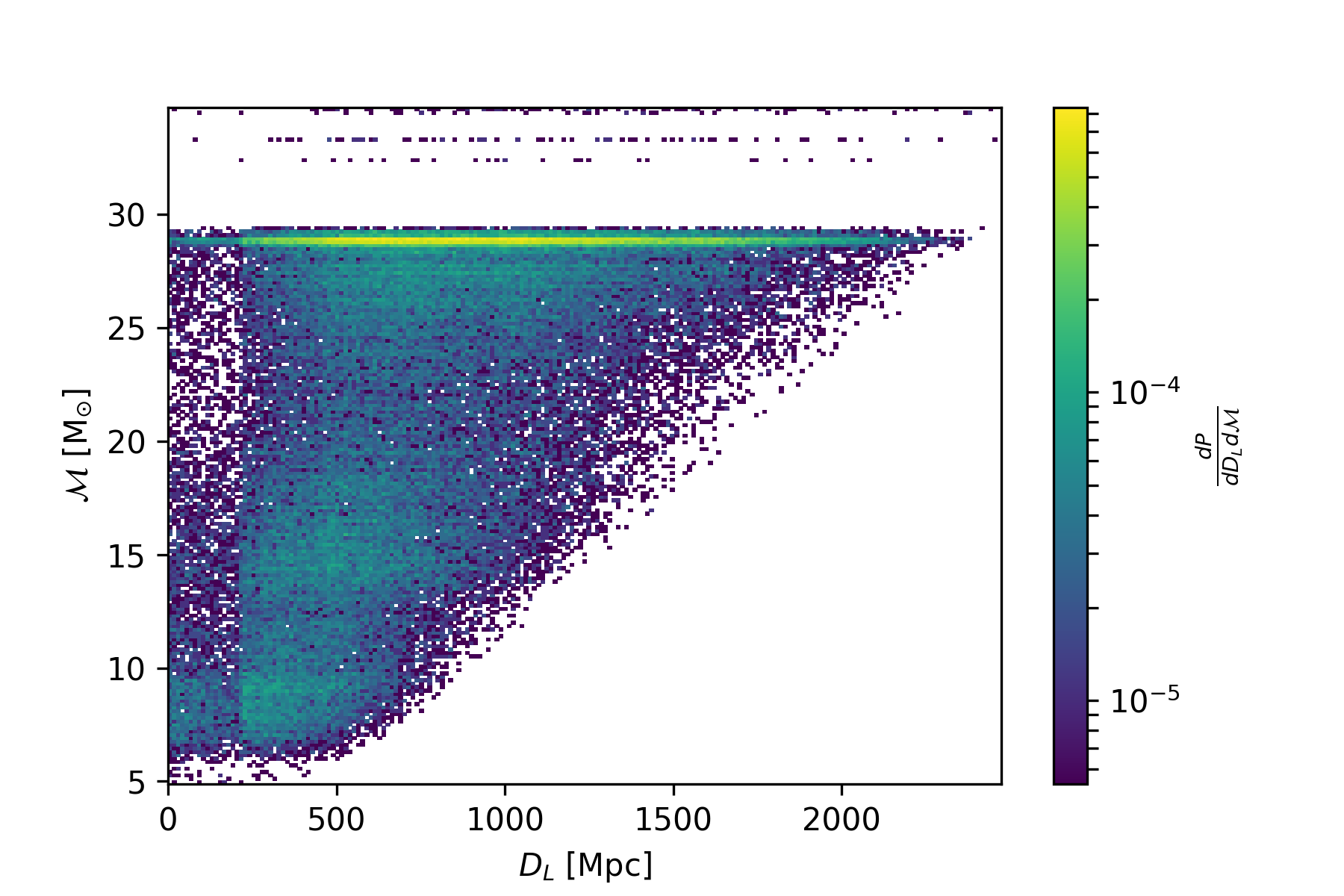}
      \caption{Chirp mass--luminosity distance distribution for Variant 12 of the \texttt{StarTrack} model with $Z=0.1Z_{\odot}$.}
         \label{fig:ABHBH.002v12}
   \end{figure}

An unexpected result is the strong anti-symmetry of the TOD for some models. We do not expect full symmetry in Tables \ref{tab:perc10} and \ref{tab:perc1} because the corresponding TODs are calculated using different observations.
Still, in some cases, breaking of this symmetry is extreme enough that we propose an explanation based on the distributions for Standard \texttt{StarTrack} model and Variant 12 with metallicity $Z=0.1Z_\odot$. If we assume that Variant 12 is correct, we can exclude the Standard model with only 28 observations (p-value = 0.9); yet the reverse is not true even for 3\,000 observations. 
The probability distribution of both models is concentrated in the area around $\mathcal{M}\approx 28 M_\odot$. For Variant 12 the probability distribution is denser in this area than for the Standard model, while the latter distribution is overall more uniform. This fact is understated in Figs. \ref{fig:ABHBH.002S} and \ref{fig:ABHBH.002v12} due to the logarithmic scale.
When generating observations using Variant 12 one quickly exceeds the expected density for the Standard model, therefore the Variant 12 is distinguishable from the Standard model.
On the other hand,  more observations must be generated
using the more uniform distribution of the Standard model in order to
see that the density in the $\mathcal{M}\approx 28 M_\odot$ area is not as high as
expected for Variant 12.

\begin{table*}
            
\caption{Thresholds of distinguishability with p-value $\alpha=0.1$}
\centering          
\begin{tabular}{l|llllllllllllllll}
\hline\hline  
- & GC & S & V1 & V2 & V3 & V5 & V6 & V7 & V8& V9 & V10 & V11 & V12 & V13 & V14 & V15 \\
\hline 
GC &-& 4   & 3  & 4  & 10  & 4    & 3   & 4   & 4  & 4   & 4   & 3 & 3   & 5  & 5   & 4  \\
S & 4 &-  & 13 & 35 & 22  & n  & n & n & 22 &n & 155 & 5 & 28  & 6  & 58  & 28 \\
V1 & 4 & 28  &- & 74 & 10  & 28   & 35  & 28  & 6  & 22  & 45  & 4 & 28  & 4  & 28  & 22 \\
V2 & 6 & 35  & 17 &- & 28  & 35   & 45  & 35  & 8  & 35  & 58  & 4 & 35  & 5  & 533 & 17 \\
V3 & 8 & 10  & 4  & 10 &-  & 10   & 13  & 13  & 8  & 13  & 13  & 4 & 8   & 13 & 22  & 6  \\
V5 & 4 & 121 & 17 & 45 & 22  &-   &n &n & 22 &n & 121 & 5 & 28  & 6  & 58  & 28 \\
V6 & 4 & 155 & 17 & 45 & 22  &n  &-  &n & 22 &n & 121 & 5 & 22  & 6  & 58  & 28 \\
V7 & 4 & 121 & 13 & 28 & 22  & 255  & 255 &-  & 28 &n & 74  & 5 & 13  & 6  & 45  & 22 \\
V8 & 3 & 8   & 3  & 3  & 8   & 8    & 8   & 8   &- & 10  & 5   & 4 & 5   & 5  & 5   & 3  \\
V9 & 4 & 95  & 10 & 22 & 22  & 121  & 121 & 199 & 35 &-  & 45  & 5 & 17  & 6  & 45  & 17 \\
V10 & 4 & 533 & 22 & 45 & 22  & 1832 & 874 & 417 & 17 & 326 &-  & 5 & 155 & 6  & 74  & 35 \\
V11 & 3 & 3   & 3  & 3  & 3   & 3    & 3   & 3   & 4  & 3   & 3   &-& 3   & 3  & 3   & 3  \\
V12 & 4 &n & 22 & 45 & 22  &n  &n &n & 17 &n & 417 & 5 &-  & 6  & 74  & 45 \\
V13 & 3 & 5   & 3  & 3  & 10  & 5    & 5   & 5   & 13 & 6   & 4   & 3 & 4   &- & 4   & 4  \\
V14 & 6 & 22  & 8  & 35 & 121 & 22   & 22  & 22  & 8  & 22  & 22  & 4 & 17  & 8  &-  & 10 \\
V15 & 4 & 95  & 28 & 22 & 10  & 95   & 95  & 95  & 17 & 74  & 121 & 5 & 45  & 5  & 22  &-
\end{tabular}
\tablefoot{Displayed vertically are the tested models. Displayed horizontally are the models assumed to be correct, i.e. used to generate data $O_i$. Here, $n$ signifies that the particular pair of models is indistinguishable even after 3\,000 observations. } 
\label{tab:perc10}  
\end{table*}

\begin{table*}
\caption{Thresholds of distinguishability with p-value $\alpha=0.01$}
\centering          
\begin{tabular}{l|llllllllllllllll}
\hline\hline  
- & GC & S & V1 & V2 & V3 & V5 & V6 & V7 & V8& V9 & V10 & V11 & V12 & V13 & V14 & V15 \\
\hline 
GC &- & 13   & 8  & 13  & 28  & 10  & 13   & 13   & 10 & 13   & 13   & 10 & 13  & 17 & 13   & 10  \\
S & 17 &-   & 58 & 155 & 74  &n &n  &n  & 74 &n  & 683  & 17 & 155 & 17 & 199  & 121 \\
V1 & 13 & 95   &- & 199 & 28  & 95  & 74   & 74   & 22 & 74   & 95   & 13 & 95  & 10 & 95   & 58  \\
V2 & 22 & 121  & 58 &-  & 95  & 155 & 155  & 95   & 28 & 121  & 199  & 13 & 95  & 13 & 1832 & 58  \\
V3 & 28 & 35   & 17 & 35  &-  & 35  & 35   & 45   & 22 & 45   & 35   & 10 & 28  & 45 & 74   & 22  \\
V5 & 17 & 417  & 58 & 155 & 95  &-  &n  &n  & 58 &n  & 417  & 17 & 95  & 17 & 199  & 121 \\
V6 & 17 & 417  & 58 & 121 & 74  &n &-   &n  & 58 &n  & 417  & 17 & 95  & 17 & 199  & 95  \\
V7 & 13 & 417  & 45 & 95  & 74  & 874 & 874  &-   & 95 &n  & 199  & 17 & 74  & 17 & 155  & 74  \\
V8 & 8  & 28   & 10 & 13  & 28  & 28  & 35   & 28   &- & 35   & 22   & 13 & 17  & 13 & 17   & 17  \\
V9 & 13 & 255  & 45 & 74  & 74  & 417 & 417  & 533  & 95 &-   & 199  & 17 & 58  & 17 & 121  & 58  \\
V10 & 17 & 1832 & 58 & 199 & 74  &n & 3000 & 1431 & 45 & 1118 &-   & 17 & 417 & 17 & 199  & 121 \\
V11 & 13 & 3    & 6  & 5   & 3   & 3   & 3    & 3    & 13 & 3    & 3    &- & 5   & 3  & 3    & 5   \\
V12 & 17 &n  & 58 & 199 & 74  &n &n  &n  & 58 &n  & 1431 & 13 &-  & 17 & 199  & 155 \\
V13 & 8  & 17   & 6  & 10  & 35  & 13  & 17   & 17   & 35 & 17   & 17   & 8  & 13  &- & 17   & 10  \\
V14 & 22 & 74   & 28 & 95  & 533 & 74  & 74   & 74   & 28 & 74   & 74   & 13 & 45  & 22 &-   & 28  \\
V15 & 17 & 326  & 95 & 74  & 35  & 417 & 326  & 255  & 58 & 255  & 326  & 13 & 199 & 17 & 58   &-    
\end{tabular}
\tablefoot{Displayed vertically are the tested model. Displayed horizontally are the models assumed to be correct, i.e. used to generate data $O_i$. Here, $n$ signifies that the particular pair of models is indistinguishable even after 3\,000 observations. }
\label{tab:perc1}      
\end{table*}

\label{ssec:thesholds}

\section{Conclusions}
Tables \ref{tab:perc10} and \ref{tab:perc1} show that given 100 observations we will likely be able to distinguish between almost 80\% of the pairs of models. Given 1000 observations we will obtain p-value$=0.01$ certainty in distinguishing between those models. Nevertheless, 16 out of the 240 pairs of models will remain indistinguishable, or rather we do not know how many observations we would need because of the imposed limit of 3\,000 observations in our analysis. 
The current number of observations allows us to distinguish between 173 out of 240 pairs of models with $\alpha=0.1$ and between 102 pairs with $\alpha=0.01$.
This threshold of 100 observations will likely be reached once the full results of the next observational run O3 are published \citep{2010CQGra..27q3001A}. 
The most likely is Variant 9, followed closely by Variants 7, 6, and 5, and also the Standard \texttt{StarTrack} model.
Several \texttt{StarTrack} models with metallicity of $Z=0.02$ are likely to explain observations from O1 and O2, but no model is capable of explaining all observations from O3. The last runs contain mergers with chirp masses and distances that are greater than what the \texttt{StarTrack} models can account for.
A possible solution is that the observed mergers followed several different channels of evolution and no single model can explain all of the observations.

\begin{acknowledgements}
Special thanks to Tomasz Bulik for indispensable discussion as well as the insights about stellar evolution physics. 
This work was supported by the TEAM/2016-3/19 grant from FNP.
\end{acknowledgements}

%-------------------------------------------------------------------
\bibliographystyle{aa}
\bibliography{referencje.bib}{} 
\end{document}